\documentclass[aps,reprint,pre,amsmath,amssymb,superscriptaddress]{revtex4-1}
\usepackage{graphicx}

\newcommand{\lu}{\lambda_\text{u}}
\newcommand{\ls}{\lambda_\text{s}}
\newcommand{\omb}{\omega_\text{b}}

\newcommand{\kT}{k_{\text B}T}
\newcommand{\vmin}{v_\text{min}}

\let\Re\relax
\DeclareMathOperator{\Re}{Re}

\begin{document}

\title{Persistence of transition state structure in chemical reactions\\ 
        driven by fields oscillating in time}
\author{Galen T. Craven}
\affiliation{Center for Computational Molecular Science and Technology, \\
School of Chemistry and Biochemistry, \\
Georgia Institute of Technology, \\
Atlanta, GA  30332-0400}

\author{Thomas Bartsch}
\affiliation{Department of Mathematical Sciences, \\
Loughborough University, \\
Loughborough LE11 3TU, \\
United Kingdom}

\author{Rigoberto Hernandez}
\thanks{Author to whom correspondence should be addressed}
\email{hernandez@chemistry.gatech.edu.}
\affiliation{Center for Computational Molecular Science and Technology, \\
School of Chemistry and Biochemistry, \\
Georgia Institute of Technology, \\
Atlanta, GA  30332-0400}

\begin{abstract}
Chemical reactions subjected to time-varying external forces
cannot generally be described through a fixed bottleneck near
the transition state barrier or dividing surface.
A naive dividing surface attached to the instantaneous,
but moving, barrier top also fails to be recrossing-free.
We construct a moving dividing surface in phase space
over a transition state trajectory.
This surface is recrossing-free for both Hamiltonian and dissipative dynamics.
This is confirmed even for strongly anharmonic barriers using simulation.
The power of transition state theory
is thereby applicable to chemical reactions and other activated processes
even when the bottlenecks are time-dependent
and move across space.
\end{abstract}

   \maketitle

A ubiquitous problem in physics concerns the determination of 
the mechanism and rate of crossing a bottleneck
from initial to final states.
In the usual cases, the bottleneck is fixed in time and
corresponds to a saddle point 
(or a ridge of the potential in dimension two or higher)
that determines the dynamics. 
These structures lose their dynamical significance if the potential is time-dependent. 
However, in those cases in which the barrier moves up and down,
perhaps even stochastically, an invariant structure 
associated with the bottleneck persists \cite{Lehmann00a,Lehmann00b,Lehmann03,Maier01,Dyk04,Dyk05}.
This is perhaps not surprising because the saddle point of 
the potential---that is, the barrier top---remains fixed. 
But what if the {\it position} associated with the barrier top
moves with time and hence the bottleneck is not fixed? 
In this Rapid Communication, we show that under some conditions---namely when the
motion of the barrier top is periodic---there still exists
a fixed structure associated with the 
bottleneck---the transition state trajectory \cite{dawn05a,dawn05b}.

This result is of particular interest to chemical physics in which
the determination of rates is a central concern, and increasingly
rates must be determined in systems that are driven far from equilibrium.
Specifically, the response of a chemical constituent to the external forcing 
by oscillating fields can strongly 
influence the mechanism and rate in which a reactant is transformed to product.
Organic polarization synthesis \cite{Lids01} and 
colloidal and macromolecular structure assembly \cite{Elsner09,Jager11,Prokop12,Ma13}
offer examples of such time-dependent chemical transformations 
driven under kinetic control. 

For field induced molecular dissociation \cite{Hiskes61},
formaldehyde (H$_2$CO) \cite{Moore83,mill90a,Butenhoff91}
can be considered as a prototypical example.
The potential energy surface (PES) of formaldehyde
contains two dissociation channels (H$_2$+CO and H+HCO)
and isomerization channels
to \textit{cis}-HCOH and \textit{trans}-HCOH isomers \cite{Townsend04,Zhang04}.
When H$_2$CO is subjected to the influence of an external laser field,
it is directionally forced.
This forcing deforms the PES and influences
the reaction rates as well as the placement of the
transition state dividing surface.
Interest in the construction of a recrossing-free dividing surface (DS) in 
the bottleneck region of phase space, where reactive trajectories must cross, 
is not confined to the field of chemical physics \cite{hern10a}.
For example, bottlenecks play an important role in the dynamics of
atoms \cite{Jaffe99},
clusters \cite{Komatsuzaki99},
microjunctions \cite{ballistic},
asteroids \cite{Jaffe02},
and cosmological spacetime models \cite{Oliveira02}.

In the absence of a driving field,
transition state theory (TST) \cite{truh83,Truhlar96,Miller,hern10a} offers
a formally exact rate calculation in chemically reactive systems.
The methodological hurdle in such calculations is the construction of 
a hypersurface in phase space 
that separates reactant and product regions and
that is crossed only once by all reactive trajectories.
If such a DS cannot be constructed, 
the TST rate is no longer exact but only an upper bound to the rate.
Indeed, variational
transition state theory 
\cite{truh84,hynes85b,pollak90a,truh2000,pollak05a,Peters14}
has been extremely effective at providing 
relatively high-accuracy approximations to the rate and the DS.
The aim of this article is to resolve the structure of the transition
state geometry in situations where the transition state is not fixed because
the driving field is oscillatory, 
advancing previous work by two of us on time-dependent TST \cite{hern08d}.

\begin{figure}[t!]
\includegraphics[width=8.0cm,clip]{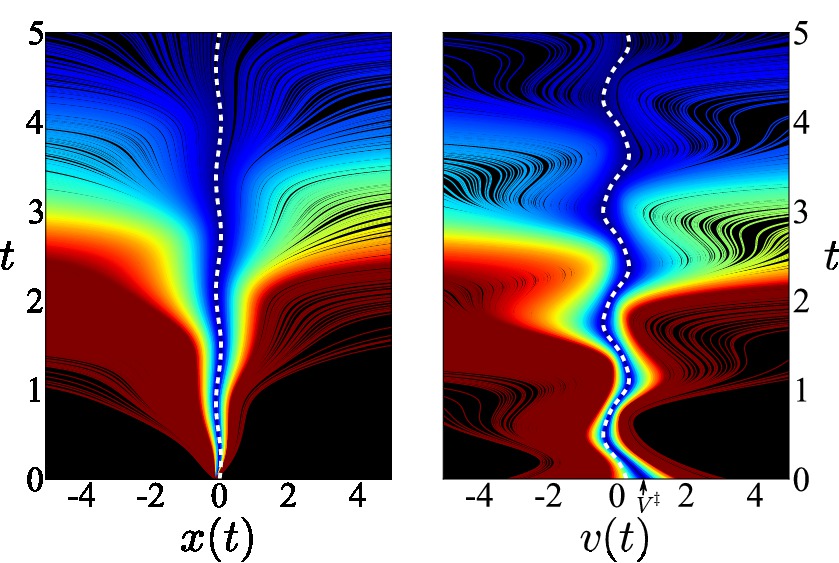}
\caption{\label{fig:x_v}
(Color online)
Time evolution of $x(t)$ (left) and $v(t)$ (right) for a swarm of 2000 trajectories following the equation of motion~(\ref{eq:motionanharm}).
The transition state trajectory (TS) is shown in dashed white. 
The trajectories are colored according to the difference in initial velocity, $\left|V^\ddag-v_0\right|$, with respect to the critical velocity $V^\ddag$
as marked. 
The colors range from dark blue to dark red 
(or black to gray in grayscale) and are scaled, increasingly, by this difference.
Parameters are $\epsilon = 2$, $\Omega=5$, $\gamma=3$, $a=1$, and $\varphi=0$.
}
\end{figure}

In an autonomous system with two degrees of freedom, 
Pechukas and Pollak have shown that the optimal dividing surface is an unstable periodic orbit (PO) \cite{pollak78,pollak79,pech79a,pollak80}. 
Its projection into configuration space provides a dividing surface
that is locally recrossing free.
In systems with three or more degrees of freedom,  this periodic orbit is 
generalized to a normally hyperbolic invariant manifold (NHIM) \cite{deLeon2,hern93b,hern94,Uzer02,hern10a,Allahem12,Li06prl,Waalkens04b,Waalkens13}. 
Attached to the NHIM are stable and unstable manifolds.
These manifolds form phase space separatrices that distinguish between 
reactive and nonreactive trajectories and also constitute the pathways 
by which reactive trajectories are funneled 
from reactant to product through the transition state \cite{Uzer02,Allahem12,Waalkens13}.
The central result of this article, elaborated below,
is that there is a sense in which this structure persists even when the
chemical reaction is driven by an external oscillating field as, for example,
from an external electric field \cite{Kawai07,Kawai11laser}.

A particle of unit mass propagating from an initial position $x_0$ and surmounting a moving one-dimensional energy barrier
serves as a paradigm for the present approach.
The barrier is moving with a time-dependent, instantaneous position $E(t)$,
and is specified by
\begin{equation}
	\label{eq:potanharm}
	U(x) = -\tfrac 12 \omb^2 (x-E(t))^2-\tfrac{1}{4}\epsilon(x-E(t))^4.
\end{equation}
It leads to the equation of motion
\begin{equation}
	\label{eq:motionanharm}
	\ddot x + \gamma \dot x  =  \omb^2 (x -  E(t))+\epsilon (x -  E(t))^3,
\end{equation} 
where $E(t)$ is a driving field, $\gamma$ is a dissipative emission parameter, $\omb$ is the barrier frequency, and $\epsilon$ is an anharmonic coefficient.
We consider here both the harmonic ($\epsilon=0$) and anharmonic
($\epsilon\ne0$) cases.
In the latter case, the reacting particle's degree of freedom is non-linearly coupled to the 
motion of the driving field.

When $\gamma=0$, the system is Hamiltonian and the dynamics are representative of a chemical reaction forced by an external field, such as a laser.
The coupling of a molecule's dipole moment with an external field is known to accelerate the dynamics of a chemical reaction.
The collinear H+H$_2$ exchange reaction is an example 
of a physical system that can be represented through (\ref{eq:motionanharm}).
The asymmetric stretch of the system creates a time-dependent dipole in the region of the one-dimensional TS.
The rate of barrier crossing is accelerated when the dipole couples with an external driving field \cite{Miller80}.

For dissipative ($\gamma>0$) systems, Eq.~(\ref{eq:motionanharm}) is a classical approximation for 
a field-induced reaction undergoing spontaneous emission along a reaction coordinate \cite{Argonov08}. 
Herein, we show that when a chemical reaction is forced by a temporally periodic external field, there persists a strictly recrossing-free DS. 
This recrossing-free criterion is satisfied even for systems that are undergoing a cooling process, i.e., $\gamma>0$.

For every $E(t)$ there exists a specific trajectory that remains close to the energy barrier for all time and never descends into either product or reactant regions. 
This trajectory has been termed the transition state trajectory (TS) \cite{dawn05a,dawn05b,hern06d,Revuelta12,Bartsch12}.
We will use a time-dependent DS that is located at the instantaneous position of the TS~trajectory and
show that this DS is recrossing-free, thus confirming that a transition state persists in non-autonomous systems. 
However, it does not correspond to the location of an energetic saddle point, i.e., an activated complex.

In the harmonic ($\epsilon=0$) case for an arbitrary driving field $E(t)$, Eq.~(\ref{eq:motionanharm})
can be solved exactly, 
with the eigenvalues $\lambda_{\text s,u} = -\frac{1}{2} \left(\gamma \pm \sqrt{\gamma^2 + 4 \omb^2}\right)$ 
corresponding to the stable and unstable manifolds.  
Particular solutions of Eq.~(\ref{eq:motionanharm}) can be expressed through the $S$ functionals~\cite{dawn05b,Kawai07}
\begin{equation}
    \label{eq:SDef}
    S_\tau[\mu, g;t] = \begin{cases}
            \displaystyle -\int_t^\infty g(\tau)\,\exp(\mu(t-\tau)) \,d\tau \!\!\!
                & :\; \Re\mu>0, \\[3ex]
            \displaystyle +\int_{-\infty}^t g(\tau)\,\exp(\mu(t-\tau)) \,d\tau \!\!\!
                & :\; \Re\mu<0,
        \end{cases}
\end{equation}
where $\mu$ is an eigenvalue of (\ref{eq:motionanharm}) 
and $g(\tau)$ is a time-dependent modulation to the autonomous intramolecular potential. 
The general solution will contain stable and unstable components, given by (\ref{eq:SDef}), 
and an exponential term which must be omitted to obtain a bounded solution \cite{dawn05b}.
The TS trajectory is therefore given by~\cite{Revuelta12,Bartsch12}
\begin{equation}
\begin{aligned}
    \label{eq:TStraj_x}
    x^\ddag(t) &= \frac{\omb^2}{\lu-\ls}\,\left(S[\ls, E;t]-S[\lu, E;t]\right), \\
    v^\ddag(t) &= \frac{\omb^2}{\lu-\ls}\,\left(\ls S[\ls, E;t]-\lu S[\lu, E;t]\right).
\end{aligned}
\end{equation}
Equation (\ref{eq:TStraj_x}) gives the TS solution for any $E(t)$,
 provided only that $\epsilon=0$ and that $E(t)$ is polynomially bounded for $t\to\pm\infty$, 
such that the $S$ functionals exist.

We now restrict the discussion to 
sinusoidally oscillating fields of the form
\begin{equation}
	\label{eq:oscBarrier}
	E(t) = a \sin(\Omega t + \varphi),
\end{equation}
although the methods presented herein apply equally to arbitrary periodic oscillations.
With this restriction, the TS trajectory, given by Eq.~(\ref{eq:TStraj_x}),
is an unstable PO whose period $2 \pi/\Omega$ is the period of the external driving.
In systems with anharmonic barriers, we will therefore choose 
an unstable PO close to the barrier top as the TS~trajectory. 
The TS trajectory acts like  a moving saddle point:
Like the equilibrium point on the autonomous barrier, it remains in the transition region for all time. 
Trajectories that begin on the stable manifold approach it, asymptotically, as $t\to\infty$. 
All other trajectories move away in the infinite future.

Fig.~\ref{fig:x_v} shows the time evolution of $x(t)$ and $v(t)$ for a set of trajectories starting at some point $x_0$ to the left of the barrier.
Specifically, the potential (\ref{eq:potanharm}) describes an inverted (an)harmonic oscillator.
Initial velocities are sampled from a Boltzmann distribution $q(v)$.
For all numerical simulations in this paper,
we have chosen units such that the particle mass, the barrier frequency $\omb$ and the thermal energy $\kT$ of the initial Boltzmann distribution are unity;
all other parameters are dimensionless.
Most trajectories in Fig.~\ref{fig:x_v} 
quickly move away from the DS in accordance with the unstable nature of the PO.
As a consequence of this instability, the Poincar\'e return map
that records the phase space position of a trajectory after each period of the driving
contains very little information
Though not shown, it has a single fixed point arising from the TS~trajectory 
and only a few returns for the escaping trajectories.

\begin{figure}
\includegraphics[width = 8.0cm,clip]{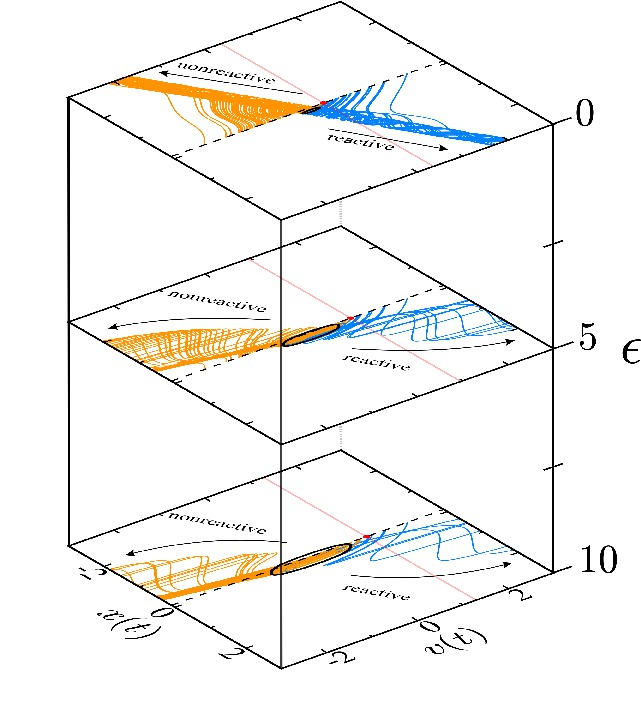}
\caption{\label{fig:manifolds}
(Color online)
A swarm of 100 trajectories starting at $x_0$ (dashed black line)
propagated by Eq.~(\ref{eq:motionanharm})
at $\epsilon=0$, 5 and 10.
Reactive and nonreactive trajectories are identified by labeled arrows.
The TS~trajectory is a periodic orbit in solid black.
The critical velocity $V^\ddag$ lies at the vertex of $x_0$ and the 
grey (red) line.
Parameters are $\Omega=5$, $\gamma=1$, $a=1$, and $\varphi=0$.
}
\end{figure}

Some trajectories, however, remain close to the TS~trajectory for long times.
Indeed, given an initial position, there exists a unique trajectory 
that approaches the TS~trajectory asymptotically with increasing time.
It can be specified by its initial velocity, 
which we call the critical velocity $V^\ddag$.
This particular trajectory lies on the stable manifold of the TS~trajectory
(which by definition contains all those trajectories that asymptotically approach the TS~trajectory as $t\to\infty$).
Trajectories close to the stable manifold are captured in the vicinity of the 
TS~trajectory for a long time
before they finally descend into either the reactant or the product wells.
The stable manifold itself contains trajectories that will never descend.
It therefore separates reactive from nonreactive trajectories in phase space:
Trajectories whose initial velocity is larger than $V^\ddag$ are reactive,
those with initial velocities below $V^\ddag$ are not.

In our numerical computation,
we choose initial conditions on the line $x=x_0=-0.1$.
The stable manifold intersects this line at the point $(x_0,V^\ddag)$.
In the present case,
the critical velocity is $V^\ddag\approx0.819$
as highlighted in Fig.~\ref{fig:x_v}.
Note that it is not the velocity of 
the instantaneous barrier top,
which is $a\Omega=5$ at $t=0$.

The TS~trajectory also defines a moving DS, $x=x^\ddag(t)$,
that can be used to track the reactant and product populations
in the generic reaction
$\text{R} \rightarrow \text{P}$.
The normalized reactant population $P_\text{R}(t)$ 
is the fraction of trajectories that are on the reactant side of the TS~trajectory,
relative to the moving DS, at time $t$. 
In a two-state model, the normalized product population 
is $P_\text{P}(t) = 1-P_\text{R}(t)$.  
A monotonic behavior in these populations
indicates that the chosen DS is recrossing-free. 

\begin{figure}[t]
\includegraphics[width = 8.0cm,clip]{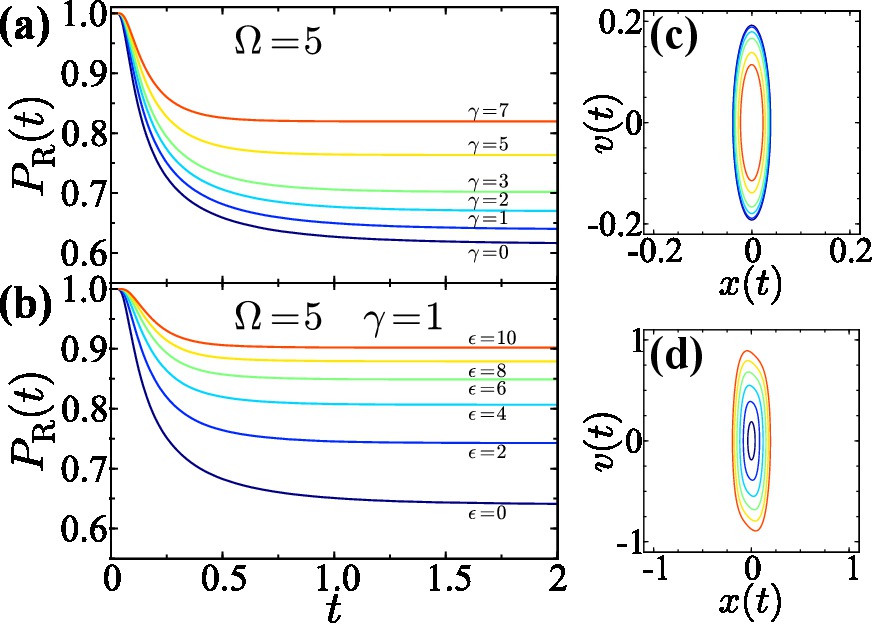}
\caption{\label{fig:Pop_Flux}
(Color online)
Reactant populations as a function of time for the 
harmonic (a) and anharmonic (b) barrier.
The corresponding TS trajectories are shown in panels
(c) and (d), respectively.
In all cases, $a=1$, and $\varphi=0$.}
\end{figure}

A reactive trajectory will cross the moving DS $x=x^\ddag(t)$ at a time
$t^\ddag(v)$ that depends on the initial velocity.
At any time $t>0$, the product region $(x(t)-x^\ddag(t))>0$, 
to the right of the moving surface, will contain all those trajectories that cross the surface at a time $t^\ddag<t$. 
These are the trajectories that have an initial velocity of at least $\vmin$, where $t^\ddag(\vmin-v^\ddag(0)) = t$.  
From this condition and the expression of $t^\ddag$ derived in Ref.~\onlinecite{hern06d},
 for a harmonic barrier, $\vmin$ can be obtained exactly and is given by
\begin{equation}
	\label{eq:vmin}
	\vmin(t) =v^\ddag(0) + \frac{\lu e^{-\lu t} - \ls e^{-\ls t}}{e^{-\lu t} - e^{-\ls t}}\,(x_0-x^\ddag(0)).
\end{equation}
The population of the reactant region at time $t$ is therefore
\begin{equation}
	\label{eq:pPop}
	P_\text{R}(t) =\int_{-\infty}^{\vmin(t)} q(v)\,dv.
\end{equation}
The critical velocity $V^\ddag$ is the long-time limit of $\vmin(t)$.
Because $V^\ddag$ is a time-invariant identifier of reactive trajectories, the reactant population in the long-time limit is 
\begin{equation}
	\label{eq:pPopinf}
	P_\text{R}(\infty) =\int_{-\infty}^{V^\ddag} q(v)\,dv,
\end{equation}
which is the fraction of trajectories that never surmount the barrier. 

\begin{figure}[t]
\includegraphics[width = 8.0cm,clip]{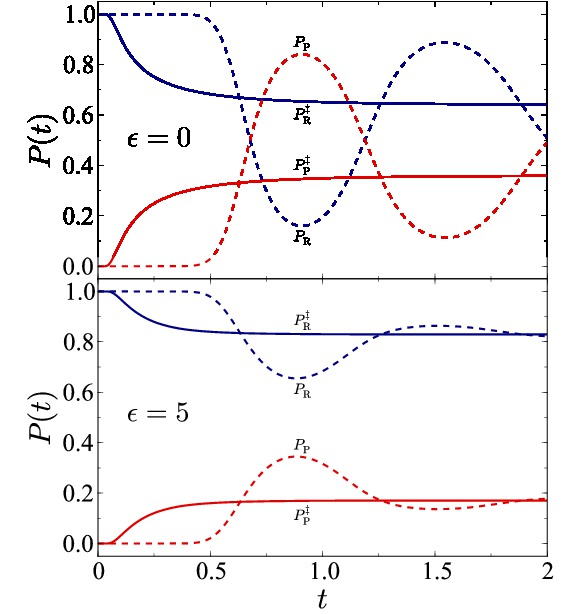}
\caption{\label{fig:recross}
(Color online)
Time dependence of the reactant $P_\text{R}$
and product $P_\text{P}$ populations
for harmonic and anharmonic barriers
obtained relative to the instantaneous barrier top (dashed lines)
and the TS~trajectory (solid lines).
All other parameters as in Fig.~\ref{fig:manifolds}.
}
\end{figure}

For an anharmonic barrier, Eqs.~(\ref{eq:pPop}), and (\ref{eq:pPopinf}) are valid, although $\vmin(t)$ is, in general, not known exactly.
Fig.~\ref{fig:manifolds} illustrates trajectories for various strengths of the anharmonicity.
The critical velocity, shown as a red circle,
marks the boundary between reactive and nonreactive trajectories.
The reactive trajectories trace the forward branch of the unstable manifold 
while the nonreactive trajectories trace the backward branch. 
The location of a trajectory's initial velocity with respect to $V^\ddag$ 
decides which branch the trajectory follows as it moves toward its final state.
It can also be seen in Fig.~\ref{fig:manifolds} that $V^\ddag$ increases with increasing $\epsilon$, 
and thus increasing the anharmonicity decreases the amount of product formed.
This increase in $V^\ddag$ is due to the curvature 
in the stable and unstable manifolds
that is induced by anharmonicity.

To test that the DS is recrossing-free, we simulated ensembles of $10^6$ trajectories
with an initial position $x_0=-0.1$ to the left of the instantaneous barrier top and initial velocities sampled from a Boltzmann distribution. 
For every time $t$ we compute the normalized reactant population $P_\text{R}(t)$
and the normalized product population $P_\text{P}(t)$.
The time evolution of $P_\text{R}(t)$ for varying parameters values is shown in Fig.~\ref{fig:Pop_Flux}. The harmonic case is shown in Fig.~\ref{fig:Pop_Flux}(a) with the corresponding TS trajectories in Fig.~\ref{fig:Pop_Flux}(c).
The anharmonic case is shown in Fig.~\ref{fig:Pop_Flux}(b) with corresponding TS trajectories shown in Fig.~\ref{fig:Pop_Flux}(d).
In all cases, the DS is free of recrossings,
as is evident from the observation that the reactant populations decrease monotonically.

This is in stark contrast to the reactant and product populations that are obtained from a DS attached to the instantaneous barrier top.
That surface can be recrossed many times.
As a consequence, reactant and product populations determined from this surface are not monotonic,
but show pronounced oscillations as a function of time,
as shown in Fig.~\ref{fig:recross}.
Using the instantaneous barrier top as a DS, in accordance with the canonical view of the transition state,
an observer would alternatingly overestimate and underestimate the reactive portion of the ensemble of initial conditions.
Populations obtained from the recrossing-free DS not only converge faster to their long-time asymptotic values,
they also approach these values monotonically
and thereby provide rigorous upper or lower bounds for the limiting values.

In summary, we have studied the dynamics of a reactant particle surmounting an oscillating energy barrier. 
A dividing surface attached to a bounded transition state trajectory has been constructed that is rigorously free from recrossing,
even when the dynamics is strongly anharmonic, strongly dissipative, or strongly driven.
In addition, whether a trajectory is reactive or not is determined by its location relative to the stable manifold of the transition state trajectory.
The knowledge of the stable manifold therefore allows prediction of the fate (reactive or nonreactive) of any trajectory,
without having to carry out a simulation.
The validity of these results has been confirmed by a 
numerical simulation of ensembles of trajectories.
The construction of this dividing surface allows for a 
formally exact TST rate calculations in periodically driven chemical reactions 
which we are pursuing in current work.

This work has been partially supported by the National Science Foundation (NSF)
through Grant No.~NSF- CHE-1112067.
Travel between partners was partially supported through the People Programme (Marie Curie Actions)
of the European Union's Seventh Framework Programme FP7/2007-2013/ under REA Grant Agreement No. 294974.

\providecommand{\noopsort}[1]{}\providecommand{\singleletter}[1]{#1}%

\newpage
\printtables
\newpage

\end{document}